\documentclass[prl,article,twocolumn,preprintnumbers,superscriptaddress,amsmath,amssymb,aps,longbibliography]{revtex4-2}
\usepackage[utf8]{inputenc}
\usepackage{slashed,cancel}
\usepackage{bm}
\usepackage{pifont}
\usepackage[english]{babel}
\usepackage[normalem]{ulem} 
\usepackage{booktabs}
\usepackage{braket}
\usepackage{MnSymbol}
\usepackage{graphicx}
\usepackage{subcaption}
\usepackage{multirow}
\usepackage[skins,theorems]{tcolorbox}
\usepackage{hyperref}
\usepackage[capitalise]{cleveref}
\usepackage{orcidlink} 
\usepackage{ragged2e}
\usepackage{float}

\tcbset{highlight math style={enhanced,
  colframe=red,colback=white,arc=0pt,boxrule=1pt}}
\newcommand{\Mpl}{M_{\rm Pl}}

\newcommand{\D}{\textnormal{d}}

\renewcommand{\d}{\text{d}}
\definecolor{rosso}{cmyk}{0,1,1,0.4}
\definecolor{rossos}{cmyk}{0,1,1,0.55}
\definecolor{rossoc}{cmyk}{0,1,1,0.2}
\definecolor{blu}{cmyk}{1,1,0,0.3}
\definecolor{blus}{cmyk}{1,1,0,0.6}
\definecolor{bluc}{cmyk}{1,1,0,0.1}
\definecolor{verde}{cmyk}{0.92,0,0.59,0.25}
\definecolor{verdec}{cmyk}{0.92,0,0.59,0.15}
\definecolor{verdes}{cmyk}{0.92,0,0.59,0.4}
\newcommand{\edit}[1]{{\color{black}{#1}}}
\newcommand{\editdos}[1]{{\color{black}{#1}}}

\graphicspath{{Draft_Plots/}}

\begin{document}
\preprint{IPPP/25/90}

\title{\Large Neutrino Constraints on Scalar–Tensor Gravity}

\author{Arturo de Giorgi \orcidlink{0000-0002-9260-5466
}}
 \email{arturo.de-giorgi@durham.ac.uk}
\affiliation{Institute for Particle Physics Phenomenology, Durham University, South Road, DH1 3LE, Durham, UK}

\author{Ivan Martinez Soler \orcidlink{0000-0002-0308-3003}}
\email{ivan.j.martinez-soler@durham.ac.uk}
\affiliation{Institute for Particle Physics Phenomenology, Durham University, South Road, DH1 3LE, Durham, UK}

\author{Sergio Sevillano Mu\~noz \orcidlink{0000-0002-9346-4431}}
\email{sergiosm@sas.upenn.edu}
\affiliation{Institute for Particle Physics Phenomenology, Durham University, South Road, DH1 3LE, Durham, UK}
\affiliation{Center for Particle Cosmology, Department of Physics and Astronomy, University of Pennsylvania, Philadelphia, Pennsylvania 19104, USA}

\begin{abstract}
In this work, we derive novel constraints on scalar–tensor theories from neutrino physics. Spatial variations of the background scalar field effectively generate density and position-dependent Standard Model masses, including neutrinos. 
Neutrinos are a unicum in the SM due to their ability both to propagate over galactic distances and to traverse dense media such as Earth. This makes them an ideal probe of the background scalar field, which can in turn alter flavour oscillations and supernova time delays. As we enter the era of precision neutrino physics, we are compelled to explore such a scenario. We derive expressions for the relevant observables and obtain new bounds on a broad class of scalar-tensor models. We finally map the bounds to popular screening mechanisms models, such as the Symmetron and Chameleon.
\end{abstract}

\maketitle

%\tableofcontents
%%%%%%%%%%%%%%%%%%%%%%%%%%%%%%%%%%%%%%%%%%%%%%%
%%%%%%%%%%%%%%%%%%%%%%%%%%%%%%%%%%%%%%%%%%%%%%%
\section{Motivation and summary}

In this work, we present novel bounds on scalar–tensor (ST) theories~\cite{Yasunori:Fujii_2003,Clifton:2011jh}. From an effective field theory~(EFT) perspective, these theories naturally arise as corrections to General Relativity~(GR) via higher-dimensional operators,
making them inherent to low-energy limits of various approaches to quantum gravity~\cite{Herranen:2014cua, Markkanen:2018bfx,Steinwachs:2011zs,Cicoli:2023opf,Smith:2025grk,Neckam:2025kip}.  Moreover, ST theories have recently attracted renewed interest due to their potential to alleviate several cosmological tensions~\cite{Sakstein:2019fmf,DiValentino:2021izs,Khoury:2025txd,SanchezLopez:2025uzw,Ballardini:2020iws,Smith:2025icl}.

Herein, we will focus on the Brans-Dicke type theory~\cite{Brans:1961sx}, where a scalar field ($\varphi$) couples non-minimally to the gravitational sector, allowing for a $\varphi$-dependent effective gravitational constant. On large distances, the impact of $\varphi$ on the metric can be assessed by its impact on the particle geodesics. Different experiments and tests have been performed to constrain the strength of these interactions, varying from cosmological~\cite{Avilez:2013dxa}, to astrophysical~\cite{Damour:1992we,Cardoso:2013opa,doneva2022scalarization, Fischer:2024eic,Menadeo:2025hgf} and laboratory scales~\cite{EotWash,Brax_2009,Merkowitz:2010kka,Burrage:2018pyg,Brax:2022olf,SevillanoMunoz:2022tfb,Argyropoulos:2023pmy,Kading:2023mdk,Upadhye:2012rc,Jaffe:2016fsh,Elder:2019yyp,PhysRevD.106.044040,Brax:2022olf,Burrage:2014oza,Elder:2016yxm,Burrage:2016rkv,Brax:2011hb,Brax:2018zfb,Upadhye:2012qu} (see Ref.~\cite{Burrage:2017qrf} for a comprehensive review). Taken together, these studies indicate that our universe is remarkably well described by GR, constraining any putative fifth forces to be at most ${\sim 10^{-5}}$ times the strength of gravity~\cite{Bertotti:2003rm, Williams:2005rv}. Despite the tight bounds, some models can still survive by means of ``screening mechanisms'', which naturally suppress long-range forces as a function of the environment's density. Some of the most studied screened scalar models include the ``\textit{Chameleon}''~\cite{Khoury:2003rn,Burrage:2016bwy}, which increases the field mass with larger densities and therefore shortens its range, and the ``\textit{Symmetron}''~\cite{Hinterbichler:2010es,Hinterbichler:2011ca}, which instead suppresses the coupling strength.

\begin{figure}
    \centering
\includegraphics[width=1.\linewidth]{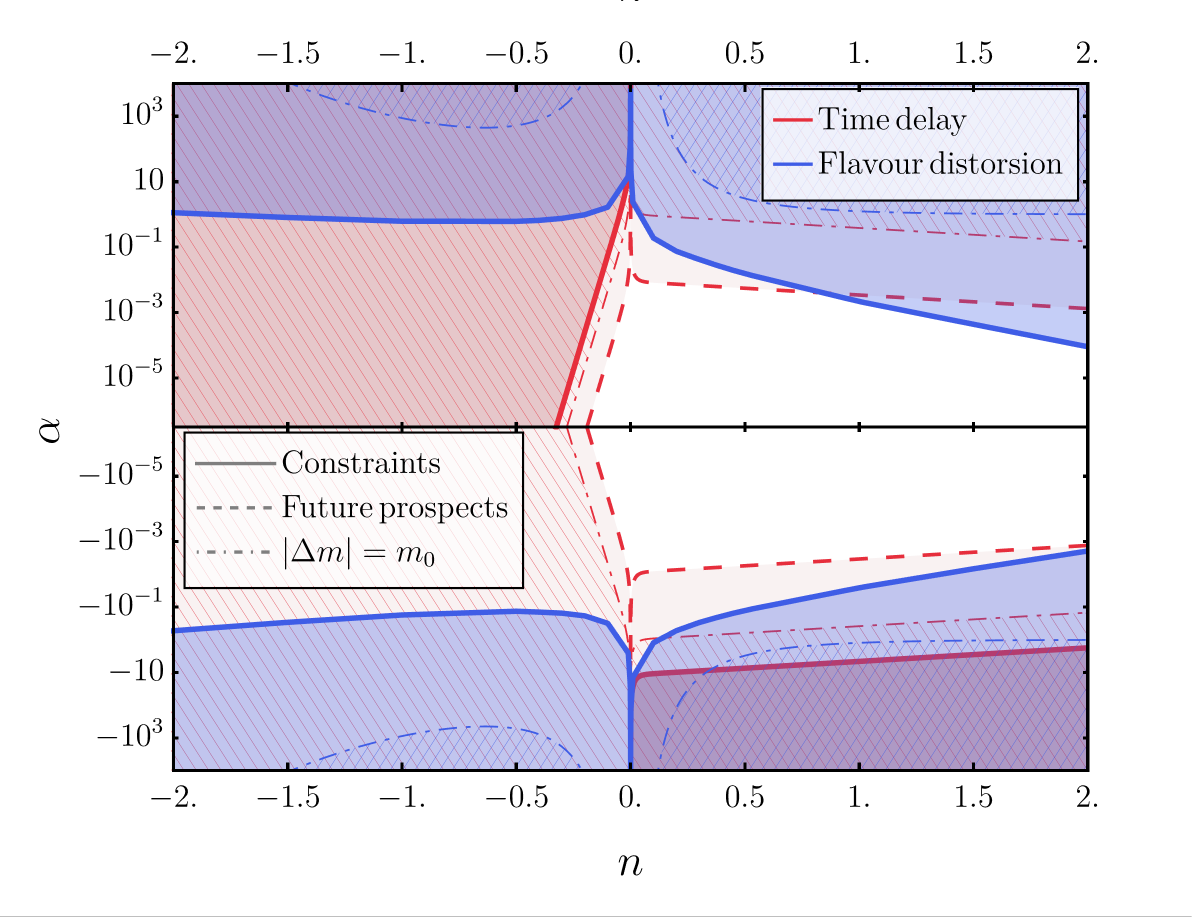}
    \caption{\justifying Summary of the region excluded in this work for the effective parameters defined in Eq.~\eqref{eq:effective}. For each analysis, we also include the EFT limit corresponding to $|\Delta m|\equiv|m_{\text{eff}}-m_0| = m_{0}$.}
    \label{fig:summary}
\end{figure}

Despite the screening, such theories can still leave important imprints on the Standard Model~(SM) as they come with a strong density-dependent vacuum expectation value~(VEV) of $\varphi$. As we will comment later on, the resulting density-dependent interactions can be equivalently interpreted as spacetime variations in SM particle masses~\cite{Burrage:2018dvt,Copeland:2021qby,SevillanoMunoz:2024ayh}. While such effects have been extremely popular in the context of time variations~\cite{Brookfield:2005bz,Copeland:2006wr,Lesgourgues:2006nd,Burrage:2024mxn}, here we test for the phenomenological implications of spatial variations~\cite{Barrow:1999qk,Levy:2024vyd,Uzan:2024ded}.
Ideally, detecting such effects requires massive particles travelling large distances and crossing regions with widely different densities, which makes the probe challenging as charged SM particles get quickly absorbed by dense media. On the other side, neutrinos, due to their extremely weak interactions, can retain information about the medium through which they propagate, making them ideal probes of the underlying scalar field configuration. This is a unicum in the SM, and can be exploited to place constraints in possibly unexplored regions of the parameter space. 
Additionally, this strategy also eliminates the need to place detectors inside unscreened regions, thereby enabling tests of environments that are otherwise inaccessible or would be screened by the presence of the experiment.

This relationship is crucial now that neutrino physics has entered the precision era. After several decades of experimental and theoretical efforts to determine the parameters that describe neutrino evolution, most of these quantities are now known to the percent level~\cite{Esteban:2024eli,Capozzi:2025ovi}.  While a few key parameters, such as the neutrino mass ordering and the CP-violation phase, remain unknown, they are expected to be determined by upcoming neutrino experiments~\cite{DUNE:2020ypp,Hyper-Kamiokande:2018ofw,JUNO:2015zny,IceCube:2025chb,KM3Net:2016zxf}. 

In this work, we derive a set of \textit{model-independent} constraints on ST theories from neutrinos. We consider a static but inhomogeneous $\varphi$ background which induces position-dependent mass-varying neutrinos. We assess the impact of neutrino observables by means of an effective phenomenological parametrisation and set bounds. In particular, we place novel bounds on these mass variations in two complementary ways: (i) by searching for deviations in the oscillation pattern, and (ii) through kinematic effects, such as changes in the time required to traverse a given distance. The study of mass-varying neutrinos through oscillation effects has been carried out in the context of scalar-dark matter-neutrino interactions~\cite{Ge:2018uhz,Choi:2019zxy,Cheek:2025kks,Sen:2023uga,Capozzi:2018bps} where the interaction lead to a modification of the neutrino masses.
The time delay in the neutrino signal caused by their mass has already been explored in the context of SN1987A observations~\cite{Zatsepin:1968kt, Loredo:2001rx, Pagliaroli:2010ik}, leading to an upper bound on the neutrino mass of $5.8$~eV. In addition to constraining the absolute neutrino mass, the time delay has also been used to probe potential interactions between dark matter and neutrinos~\cite{Ge:2024ftz}. \editdos{Here, we will use these limits to constrain scalar-tensor theories by comparing the absolute neutrino-mass scale inferred from SN1987A with independent terrestrial bounds, thereby constraining any environmental mass rescaling between Earth and interstellar space.} This bound could be lowered below the eV scale in the case of a future galactic supernova~\cite{Nardi:2003pr, Nardi:2004zg, Lu:2014zma, Hansen:2019giq, Pompa:2022cxc, Pompa:2023yzg, Parker:2023cos, Denton:2024mlb}.  A summary of our findings can be found in Fig.~\ref{fig:summary}. For the sake of concreteness, we will then apply them to the Symmetron and the Chameleon. 

%%%%%%%%%%%%%%%%%%%%%%%%%%%%%%%%%%%%%%%%%%%%%
%%%%%%%%%%%%%%%%%%%%%%%%%%%%%%%%%%%%%%%%%%%%%
\section{The Model}
\label{sec:model}
The action of the gravitational sector, including the new scalar field $\varphi$ is:
\begin{equation}\label{eq:BDgeneric}
	S_G=\int \D^4 x \sqrt{-\tilde{g}}\left[\frac{\Mpl^2 \tilde{A}(\phi)^{-2}}{2}\tilde{R} +\frac{1}{2} (\partial_\mu \phi)^2  - \tilde{V}(\phi) \right],
\end{equation}
where $\tilde{R}$ is the Ricci scalar \edit{built from the modified metric $\tilde{g}_{\mu\nu}$,} and $\tilde{V}(\phi)$ and $\tilde{A}(\phi)$ are the potential and the non-minimal coupling of $\phi$ to GR, respectively. The above action defines the so-called ``\textit{Jordan frame}''. 

The non-minimal coupling modifies gravity's dynamics and can thus be tested by studying gravitational probes. Alternatively, one can work in a different frame, where the action matches GR by performing the conformal transformation
\begin{align}
\label{eq:conformal}\tilde{g}_{\mu\nu}=\tilde{A}(\phi)^2 g_{\mu\nu}\approx \tilde{A}(\phi)^2\eta_{\mu\nu}\,,
\end{align}
where the last approximation holds if back-reaction effects of $\phi$ on the metric are subdominant. In doing so, the transformation induces couplings of $\phi$ to the SM fields via the metric. The transformed action can be written as
\begin{equation}\label{eq:BDgenericEF}
\begin{split}
	S=&\int \D^4 x \sqrt{-{g}}\left[\frac{\Mpl^2}{2}{R} +\frac{1}{2} (\partial_\mu \varphi)^2  - {V}(\varphi) \right],\\
    &+S_m[A^{2}(\varphi)g_{\mu\nu},\psi],
\end{split}
\end{equation}
where the gravitational sector only depends on the transformed metric $g_{\mu\nu}$ while the matter sector now has explicit couplings to the extra scalar field. We have further canonically normalised $\phi$ into $\varphi$, and defined $V(\varphi)\equiv \tilde{A}^4(\phi)\tilde{V}(\phi)$ and $A(\varphi)\equiv\tilde{A}(\phi)$. However, further field redefinitions
allow partial reabsorption of the $\varphi$ dependence, with exceptions of all terms which break conformal invariance~\cite{Garcia-Bellido:2011kqb,Ferreira:2016kxi,Burrage:2018dvt}. These include the kinetic term of the Higgs (which will not be relevant for our discussion), and all dimensionful quantities.
All in all, in the SM this implies that the Higgs VEV, $v$, gets rescaled by $A(\varphi)$ (see e.g. Ref.~\cite{SevillanoMunoz:2024ayh}). In this frame, hereafter referred to as ``\textit{Einstein frame}'', gravity is described by the Einstein-Hilbert action, but all SM particles feature background-dependent masses
\begin{equation}
    m_\text{eff.}(\varphi)=m_0 \times A(\varphi)\,,
\end{equation}
where $m_0$ is the mass in the Jordan frame. Within the context of neutrino physics, the same fate applies to hypothetical Majorana mass terms, which could be included alongside the right-handed neutrinos. In a more agnostic way, employing the effective field theory language, one can show that the Weinberg operator~\cite{Weinberg:1980bf} receives a $A^{-1}(\varphi)$ being a dimension-5 operator. When the Higgs gets a VEV, two extra powers of $A(\varphi)$ generate a neutrino mass term which scales with $A(\varphi)$, as previously claimed.

It is important to stress that the two frames must lead to the same physical observables. In practice, this can be used as an advantage to assess the consistency of the results and to simplify calculations, which may be more convenient in one frame rather than another. 
A summary of conversions of some quantities between the two frames that will be useful for this work is
\begin{align}
        \label{eq:frame1}&m_E= A(x)~m_J\,,
        &&n_E=A(x)^3~n_J\,,\\
         \label{eq:frame2}&G_{F,E}=A(x)^{-2}~G_{F,J}\,,
        &&u^0_E=A(x)~u^0_J\,,
\end{align}
where the labels $E,J$ labels the frames in which they are defined, $m$ is a generic mass parameter, $n$ is a number density, $G_F$ is the Fermi constant and $u^0$ is the time component of the four-velocity. Notice that, with some abuse of notation, we write $A(\varphi)\equiv A(\varphi(x)) \equiv A(x)$ to stress the most relevant dependence depending on the context.

%%%%%%%%%%%%%%%%%%%%%%%%%%%%%%%%%%%%%%%%%%%%%
%%%%%%%%%%%%%%%%%%%%%%%%%%%%%%%%%%%%%%%%%%%%%
\section{Analysis}
\label{sec:analysis}

We begin by noticing that it is always possible to redefine space-time coordinates such that at a given point $x^\mu_0$, $g_{\mu\nu}(x_0)=\eta_{\mu\nu}$, i.e. $A(x_0)=1$. The choice of $x_0$ is arbitrary, and thus unphysical. 
We choose it to lie on Earth's surface, where most of the measurements are taken.
From a phenomenological perspective, the function $A(\varphi)$ can then be expanded in a Taylor series
\begin{align}
    A(\varphi(x))&=\left(1+\sum\limits_{k=1}^\infty c_k \left(\frac{\varphi}{\Lambda_\text{cut}}\right)^k\right)/\left(1+\sum\limits_{k=1}^\infty c_k \left(\frac{\varphi_0}{\Lambda_\text{cut}}\right)^k\right)\,,\nonumber\\
    &\approx1+\sum\limits_{k=1}^\infty c_k \left[\left(\frac{\varphi}{\Lambda_\text{cut}}\right)^k-\left(\frac{\varphi_0}{\Lambda_\text{cut}}\right)^k\right]\,,
\end{align}
where the cut-off of the theory $\Lambda_\text{cut}$ and the coefficients $c_k$ depend on the model. In practice, provided the validity of the EFT, the expansion can be truncated at the leading non-vanishing order, whose order $k$ varies depending on the chosen model. Assuming the value of $\varphi$ scales with a power of the density, we can ultimately parametrise $ A(\varphi(x))$ by
\begin{equation}
       \label{eq:effective} A(\varphi(x))\approx 1+\alpha_n\left[\left(\frac{\rho(x)}{\text{g}/\text{cm}^3}\right)^n-\left(\frac{\rho_0(x)}{\text{g}/\text{cm}^3}\right)^n\right]\,,
\end{equation}
where $\rho(x)$ is the local (dark) matter density, and $(\alpha_n,n)$ are model-dependent coefficients which ought to capture the leading order correction stemming from $\varphi$. \editdos{We adopt this parametrization to ensure that no effects are present at the Earth's surface ($\rho_{0}$).} 
All the quantities with subscript $0$ are taken to be at Earth's surface, where all experiments are performed. We take $\rho_0 = 2.6~\text{g/cm}^3$, as given by the outermost layer in the PREM model~\cite{Dziewonski:1981xy}~\footnote{We also considered alternative values for $\rho_0$, such as the density of water, and found similar allowed regions for $A(\varphi)$.}.
Finally, to ensure consistency of the EFT calculations, we will always require $|A(\varphi)-1|<1$ along the neutrino trajectory.\\

%^^^^^^^^^^^^^^^^^^^^^^^^^^^^^^^^^^^^^^^^^^^^
\textbf{Flavour Distorsion --}
No significant effects are expected at the Earth’s surface as $\rho\sim\rho_0$; therefore, we do not anticipate any deviations in flavour oscillations measured by reactor or long-baseline experiments. The most promising way to probe these effects is by studying flavour oscillations as neutrinos traverse regions of matter where gravitational effects change. 

From now on, we focus on analysing flavour oscillations measured with atmospheric neutrinos. The impact of curved spacetime backgrounds on flavour oscillations has been studied extensively in the literature, beginning with the seminal works of Refs.~\cite{Piriz:1996mu,Cardall:1996cd,Fornengo:1996ef}. We build upon these results and apply them to the scenario of interest.
The evolution of the flavour eigenstates evaluated in the Jordan frame follows (cf.~\textit{Supplemental Material})
\begin{equation}
\label{eq:amplitude}
    |\psi_\alpha(s)\rangle=\exp\left(\frac{i}{2E}\int_0^s A(r)^2 H_{\alpha\beta}(r)  \,\d r\right)|\psi_\beta\rangle\,,
\end{equation}
where we have $H_{\alpha\beta}(r) = U_{\alpha k} M_{f,kj}^2U^{\ast}_{\beta j}+2E\, V_\text{CC}$ and  $V_\text{CC}=\sqrt{2}n_e G_F \delta_{e\alpha}$, with $E$ being the measured neutrino energy, $n_e$ being the electron's number density and $G_F$ the Fermi's constant. \edit{This result is frame independent, given the exponent's scale invariance. In the Jordan frame, the factor $A^2(r)$ arises from propagation in the modified spacetime metric, with one factor coming from the line element and the other from the four-momentum. In the Einstein frame, instead, the metric is canonical, while the neutrino masses depend on $A(r)$, yielding the same exponent.} This change in the oscillation phase can be understood as a consequence of neutrinos travelling through a medium in which both the effective path length and the oscillation length vary along the trajectory.
The value of $n_e$ can be inferred from Earth's mass density, which is in turn inferred by a combination of measurements of time delays of longitudinal and transverse seismic waves~\cite{Dziewonski:1981xy}.
The speed of such waves in a medium depends on the density and properties of the material, defined by its bulk and shear moduli~\cite{Kittel2005}. In the following, we will briefly comment on the case of longitudinal waves, while an analogous discussion holds for transverse ones. Denoted by $B$ the bulk modulus and by $\rho$ the density, the longitudinal wave velocity reads
\begin{equation}
    v_{L}^2 =\frac{B}{\rho}\,.
\end{equation}
Since the inference of Earth's local density is made assuming a flat metric, this implies the parameters of the above formula have to be understood as being in the Einstein frame, which can then be converted into the Jordan frame, if needed. However, this leaves an open question: is the bulk modulus also affected by the background value of $\varphi$? If so, the inferred density would rather be some effective density given by $\rho_E(x)$ weighted by additional unknown $X$ powers of $A(x)$, $\rho_\text{eff.}=\rho_E\times A^X$. 
We therefore ought to estimate the impact of $\varphi$ on the mechanical properties of the medium. A derivation of the bulk modulus from first principles is, in general, a very difficult problem, and, to the best of our knowledge, no general analytical formula exists. However, an estimation of its scaling with microscopic parameters can be obtained using the following argument:
\begin{widetext}
\begin{align}
    B&=V_0\left.\frac{\partial^2 u}{\partial V^2}\right|_{V=V_0}\sim \frac{\text{Characteristic bond energy per atom}}{\text{Characteristic atomic volume}}\sim \frac{E_\text{Rydberg}}{(R_\text{Bohr})^3}=\frac{1}{2}\left(\frac{c^5}{\hbar^3}\right)(m_e^4 \alpha^5)\propto m_e^4\,,
\end{align}
\end{widetext}
where $u$ is the interatomic potential energy per atom, $V_0$ is the material's equilibrium volume, $E_\text{Rydberg}$ and $R_\text{Bohr}$ are the Rydberg energy and the Bohr radius, respectively. The above expression reasonably captures the order of magnitude of the value of $B$ when compared with real data. Denoted by $B_0$ the measured bulk modulus on the surface, and taking into account the non-trivial $m_e(x)$ dependence in the Einstein frame, the measured quantity is then
$v_L^2=B_0 A(x)^4/\rho_E(x)$.

Therefore, the effective density accessible via measurements of sound wave velocities is
\begin{equation}
    \rho_\text{eff.}(x)=\rho_E(x)A(x)^{-4}=\rho_J(x)\,,
\end{equation}
i.e., despite the calculation being carried out in the Einstein frame, we find that the inferred numerical value of the local density coincides with the Jordan frame one.

We analysed eight years of IceCube DeepCore data~\cite{IceCubeCollaboration:2023wtb} following the parametrisation given in Eq.~\eqref{eq:effective}. At the 95\% confidence level, we found no preference for ST theories. The corresponding bound is shown in Fig.~\ref{fig:summary}. A more detailed discussion of the analysis and results is provided in the \textit{Supplemental Material}.\\

Unlike Earth, in the case of the Sun or neutrino evolution through a supernova, the evolution is adiabatic~\cite{Mikheyev:1985zog,Dighe:1999bi,deGouvea:2022dtw}. For ST theories, as shown in Eq.~\ref{eq:amplitude}, the modification in the oscillation phase arises from the factor $A(r)^2$, which alters both the mass term and the matter potential in the same way. This ensures that the evolution remains adiabatic, leading to the same mixing angles as in the standard scenario without modified gravity. Therefore, we expect no effect on flavour oscillation as the neutrino traverses these systems. See the (cf.~\textit{Supplemental Material}) for a more detailed discussion.  The impact of ST in the context of flavour oscillations through the Sun, including the Schwarzschild background, has also been explored in Refs.~\cite{MohseniSadjadi:2020xmw,Ahmadabadi:2021yfk,YazdaniAhmadabadi:2022low}.
\vspace{0.38cm}

%^^^^^^^^^^^^^^^^^^^^^^^^^^^^^^^^^^^^^^^^^^^^
\textbf{Time Delay from Supernova --}
In the scenario considered in this work, we assume that the neutrino mass is only modified after the neutrinos escape from the SN and propagate through the interstellar vacuum. As the neutrino mass changes, the time needed for the neutrinos to arrive from the supernova also does so. \edit{Solving the geodesic equation (see \textit{Supplemental Material}), the travelling time for a relativistic neutrino from the source to the lab is given by
\begin{align}
    \Delta t_\nu&=\int\limits_{r_S}^{r_0}\frac{E~dr}{\sqrt{E^2-m_0^2A(r)^2}}\approx \Delta r+\frac{m_0^2}{2E^2}\int\limits_{r_S}^{r_0}drA(r)^2\,,
\end{align}
where $E$ is the energy measured in the lab, and $r_{S,0}$ are the initial (supernova) and final (lab) positions, respectively. 
%obtained via the light-travel time, which is unaffected by conformal couplings. 
We can set constraints on $m_0A(r)$ by measuring the relative delay between} two neutrinos with \edit{same} mass but different energies, $E_{1,2}$, which reads \footnote{\editdos{We assume that neutrino propagation speeds are otherwise standard.  Any effects from additional new physics, such as Lorentz-violating dispersion or other neutrino-sector propagation effects, would be partially degenerate with the time-delay signal considered here.}}
\begin{align}\label{eq:time-delay}
    \Delta t_{21}\approx \frac{m_\edit{0}^2}{2}\left(\frac{1}{E_2^2}-\frac{1}{E_1^2}\right)\int\limits_{r_0}^{r_{S}}dr\,A(r)^2\,.
\end{align}
\edit{Here, the energy $E_{1,2}$ is a conserved quantity that is measured in the lab frame.} The correction to the SM prediction is multiplicative and acts as an effective distance. \edit{Therefore, we can constrain scalar–tensor theories by comparing the existing upper bound on the neutrino mass derived from SN1987A time-delay observations with independent local laboratory bounds on the neutrino mass, thereby limiting any environment-dependent mass shift between interstellar space and Earth.}\edit{ For $\varphi$ to be sensitive to Earth's density inside the lab it must create a thin shell, meaning that its Compton wavelength must be smaller than Earth's radius. Given such a small interaction scale, neutrinos travelling from a supernova will have a very small probability of interacting with astrophysical objects~\cite{Burrage:2023eol,March:2024oll}. However, they will still be sensitive to the screening due to dark matter, which we take to behave as a smooth dust-like fluid.} \editdos{This is a conservative assumption, given that if the scalar Compton wavelength is sufficiently short to resolve dark-matter substructure, most of the path would sample densities below the smooth-halo average, increasing the contrast with Earth and therefore generally strengthening the time-delay constraint}~\footnote{These results depend sensitively on the distribution of dark matter around galaxies. In our analysis, we model dark matter as a homogeneous fluid. If, instead, dark matter were composed of compact objects, the constraints could become significantly stronger because most of the trajectory would sample extremely low ambient densities that would average together with the high densities inside the compact objects when estimating $A(\varphi)$.}. For these results, we will use the average dark matter density in the interstellar medium along the line of sight between Earth and SN1987A. The supernova is located in the Large Magellanic Cloud~(LMC), a satellite galaxy at an angle of roughly 25 to 35 degrees below the Galactic plane as viewed from Earth~\cite{2013AJ....146...86T}. Because of its position, the line of sight avoids the dense stellar disk and instead passes primarily through the diffuse outer regions of the Milky Way NFW-halo, with the remaining portion of the path lying inside the halo of the LMC. \edit{In summary, we then} compare the mass of the neutrinos on Earth, with density $\rho_0=2.6\,\text{g/cm}^3$, to the interstellar vacuum, for which we will conservatively assume $\rho_\infty=5\times10^{-24}\,\text{g/cm}^3$~\cite{Kuhlen:2013tra}.  Given an upper bound on the mass of neutrinos from supernovae, $M_{\rm lim}$, \edit{we use Eq.~\eqref{eq:time-delay} to compare the time delay given a mass rescaling in the interstellar medium, $A_\infty$, with the one given by the laboratory frame rescaling $A_0=1$. We} take a mass rescaling $A_\infty$ to be excluded if \edit{the absolute difference is larger than the sensitivity from the existing bound, corresponding to a mass $M_{\lim}$, such that}
\begin{equation}
    |A_\infty^2m_0^2-A_0^2m_0^2|>M_{\lim}^2.
\end{equation}
The analysis of the relative timing between the arrival of events from SN1987A and their energies has allowed setting an upper bound on the neutrino mass of $5.8~\text{eV}$ at $95\%$ confidence level~\cite{Loredo:2001rx,Pagliaroli:2010ik}, but future supernovae and detectors can get this limit down to $M_{\lim}\approx0.06~$eV~\cite{Parker:2023cos}. The absolute scale of neutrino masses has not yet been measured. The current upper bound, provided by KATRIN~\cite{KATRIN:2024cdt}, is $m_0 < 0.45\,\text{eV}$. We adopt this bound as the upper limit in our analysis.
The results of this test \edit{for a generic $\rho_\infty$ dependence} are shown in red in Figure~\ref{fig:summary}\edit{, where we have used Eq.~\eqref{eq:effective}. In the next section, we will specifically calculate these bounds for two screening models.}
%%%%%%%%%%%%%%%%%%%%%%%%%%%%%%%%%%%%%%%%%%%%%
%%%%%%%%%%%%%%%%%%%%%%%%%%%%%%%%%%%%%%%%%%%%%
\section{Symmetron and Chameleon}
\label{sec:screening}
The parametrisation used in Eq.~\eqref{eq:effective} correctly captures the impact of several scalar–tensor theories featuring screening mechanisms. Herein, we will apply our results to constrain two of the most popular models: the Chameleon and the Symmetron. In the Chameleon scenario, the scalar field acquires a larger effective mass in dense environments, making the resulting interaction short-ranged and challenging to observe. 
\begin{figure*}[htb]
    \centering
    \begin{subfigure}[t]{0.487\textwidth}
    \centering
    \includegraphics[width=\linewidth]{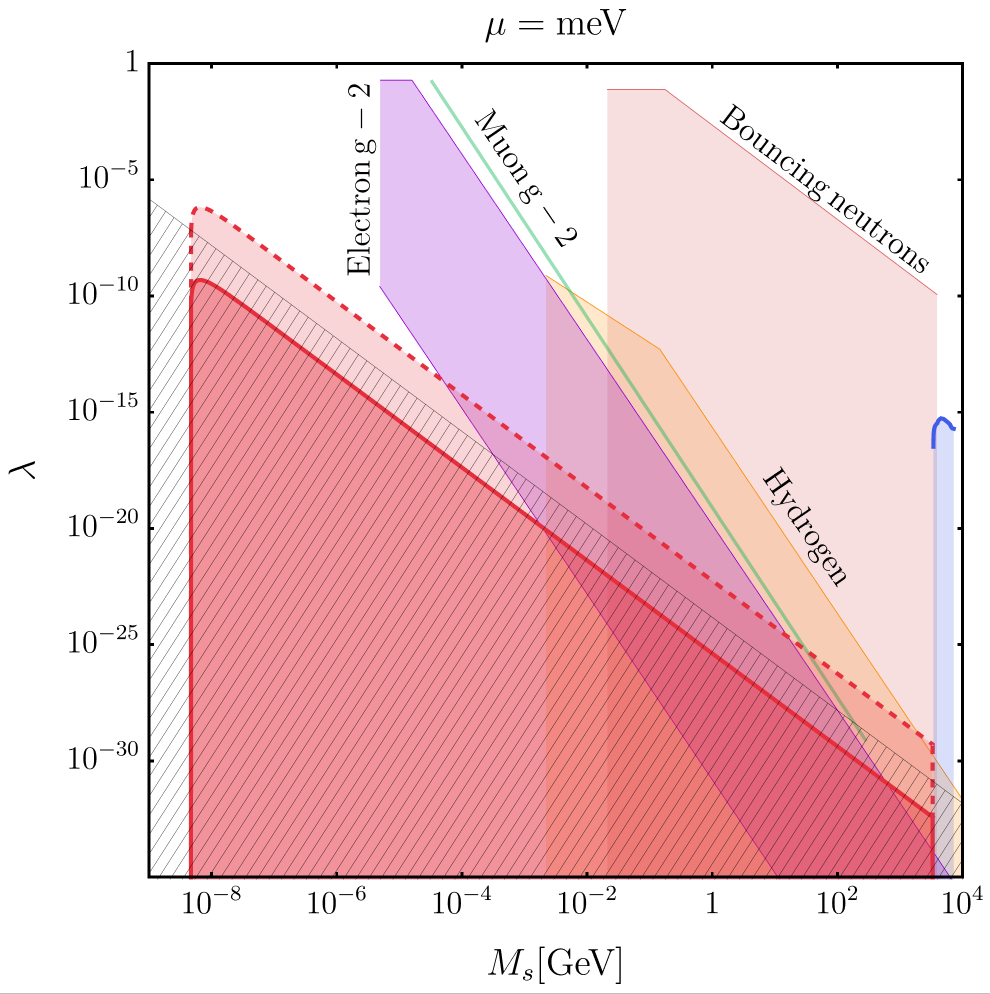}
    \caption{Symmetron}
    \label{fig:symmetron}
    \end{subfigure}
    \hfill
    \begin{subfigure}[t]{0.47\textwidth}
    \centering
    \includegraphics[width=\linewidth]{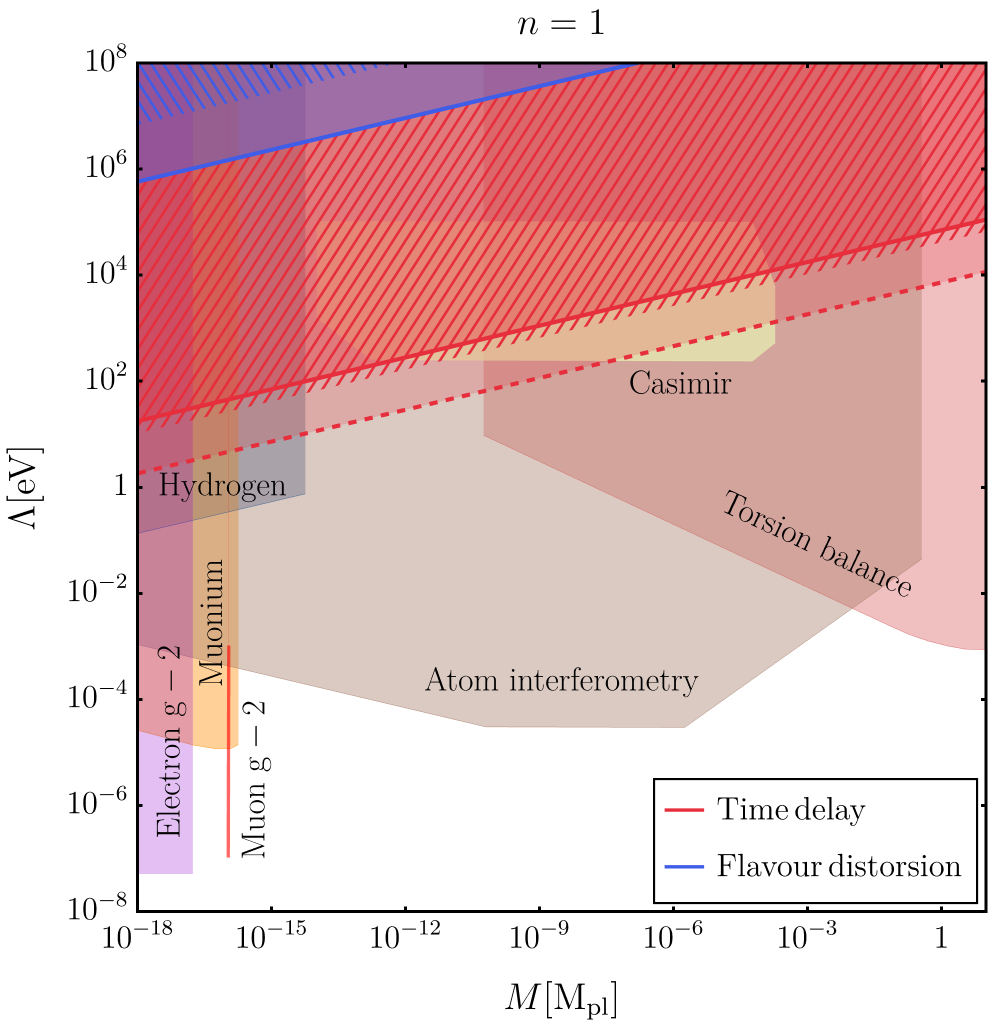}
    \caption{Chameleon}
    \label{fig:chameleon}
    \end{subfigure}
    \caption{\justifying Constraints at $95\%$~C.L. on the parameters of the Symmetron and of the Chameleon. EFT limits are represented by crossed fillings, and future prospects by dashed lines. }
    \label{fig:screening}
\end{figure*}
%^^^^^^^^^^^^^^^^^^^^^^^^^^^^^^^^^^^^^^^^^^^^

We can recast the results from Figure~\ref{fig:summary} to place constraints on these models. The conformal factors for the Chameleon and Symmetron, $A_{c,s}$, given some background density $\rho$ are defined by
\begin{align}
    &A_{\text{c},n}(\varphi)\approx1+ \left(\frac{n\Lambda^{4+n}}{M_c^n\rho}\right)^{1/(n+1)}-\left(\frac{n\Lambda^{4+n}}{M_c^n\rho_0}\right)^{1/(n+1)}\,,\\
    % &A_\text{s}(\varphi) \approx 1+\mu^2\begin{cases}
    %     \frac{1-\rho/\rho_\text{crit}}{2\lambda M_s^2}\,, & \rho \leq \rho_\text{crit}\,,\\
    %     0\,, & \rho >\rho_\text{crit} \,,
    % \end{cases}\label{eq:A_sym}\\
    % \red{+}& \red{\mu^2\begin{cases}
    %     \frac{1-\rho_0/\rho_\text{crit}}{2\lambda M_s^2}\,, & \rho_0 \leq \rho_\text{crit}\,,\\
    %     0\,, & \rho_0 >\rho_\text{crit} \,,
    % \end{cases}}\\
    &\label{eq:A_sym}A_\text{s}(\varphi) \approx 1+\frac{\mu^2}{2\lambda M_s^2}\left[
        \left(1-\frac{\rho}{\rho_\text{crit}}\right)\Theta(\rho_\text{crit}-\rho )
        \right.\\
        &\qquad\qquad\qquad\qquad\qquad\qquad\nonumber \left.-\left(1-\frac{\rho_0}{\rho_\text{crit}}\right)\Theta(\rho_\text{crit}-\rho_0 )\right]\,,
\end{align}
where $\rho_\text{crit}\equiv(\mu M_s)^2$, $\rho_0$ is the local density, $\Theta(\cdot)$ is the Heaviside step function and the residual $n$ dependence of the Chameleon comes from the choice of the potential, $V_c\propto \varphi^{-n}$. We can see that the rescalings satisfy $A(x_0)=1$
and they can be matched to the effective parametrisation of Eq.~\eqref{eq:effective} by setting
\begin{align}
        &\alpha_{\text{c},-1/(n+1)}=\left(\frac{n\Lambda^{4+n}}{M_c^n}\frac{\text{cm}^3}{\text{g}}\right)^{1/(n+1)}\,, &\alpha_{\text{s},0}=\frac{\mu^2}{2\lambda M_s^2}\,,
\end{align}
\edit{where the value for the Symmetron only applies if the density is sufficiently small such that $\rho/\rho_\text{crit}\ll 1$. Therefore, given that in this section we consider densities well beyond this limit, we derive the bounds directly using Eq.~\eqref{eq:A_sym}, rather than by recasting the results shown in Fig.~\ref{fig:summary}.}
The general constraints for the Symmetron and Chameleon can be seen in Fig.~\ref{fig:screening} for a representative choice of parameters. Note that we assumed  $A(\varphi)$ to follow the local density exactly, whereas a full scalar–tensor treatment requires solving the Klein–Gordon equation for the field profile. Our constraints therefore apply only when the scalar \editdos{closely follows the density distribution of} Earth and its core, which holds provided its Compton wavelength lies between atomic and planetary scales, corresponding to $10^{-10}<m_\varphi/\text{eV}<10^3$.

For the Symmetron, we focus on masses of $\mu=\text{meV}$, which keeps the range of the interaction within Earth's size. We find that the two types of constraints are complementary, and the range of the constrained region in each case arises from the piecewise behaviour of $A_{\rm s}$ in Eq.~\eqref{eq:A_sym}. Constraints appear only if there is a mass shift between the propagation environment and the laboratory; consequently, the critical density $\rho_{\rm crit}$ must lie between the extreme densities encountered along the line of sight. This requirement spans several more orders of magnitude in a supernova environment than in flavour oscillations within the Earth, thereby producing a correspondingly broader constrained region. Flavour oscillations can set new constraints on an area of the parameter space which has so far been unexplored. On the other hand, time delay constraints are currently not able to populate regions within the EFT validity, but can nevertheless cover a range of parameters which current bounds could not cover. Projections for the detection of the next supernova will instead be able to do so.
Larger values of $M_s$ can be probed in environments with larger densities compared to Earth as
\begin{equation}
    M_s\approx 5\times10^{3}~\text{GeV}\times \left(\frac{\rho_\text{crit}}{5~\text{g}/\text{cm}^3}\right)^{1/2}\left(\frac{\text{meV}}{\mu}\right)\,,
\end{equation}
while smaller values could be tested in lower-density regions, for example, by using extragalactic neutrinos that propagate through extremely dilute dark-matter backgrounds.

The Chameleon constraints are subdominant, as expected, since this model is already tightly tested. This is because its interaction to matter is well described by a Yukawa-type coupling across the full parameter space, unlike the Symmetron, where screening suppresses any odd term in the potential. Thus, although the Chameleon force range shrinks in terrestrial environments, subatomic laboratory experiments can still probe below these scales, producing most of the bounds shown in Figure~\ref{fig:screening}.

This illustrates the complementarity of our approach. Neutrinos feel the influence of all potential gradients along their entire trajectory, meaning they can detect short-range forces inaccessible to current laboratory tests. As a result, our method yields dominant constraints for the Symmetron’s quadratic interactions and can likewise provide strong bounds on even steeper potentials, such as $\varphi^{2n}$ couplings, which are otherwise difficult to constrain.
%%%%%%%%%%%%%%%%%%%%%%%%%%%%%%%%%%%%%%%%%%%%%
%%%%%%%%%%%%%%%%%%%%%%%%%%%%%%%%%%%%%%%%%%%%%
\section{Summary and Prospects}
\label{sec:conclusions}

In this work, we have assessed the constraining power of neutrino physics on scalar-tensor theories~(ST) and obtained new bounds. We considered the impact on neutrino flavour oscillation through Earth and time delay effects from the supernova SN1987A.  We first derived model-independent constraints based on an effective model, and then applied them to the Symmetron and Chameleon models. 

Neutrinos provide bounds that are complementary to existing ones. The long distance travelled by supernova neutrinos allow us to probe small Symmetron mass scale ($M_{s} \leq 10^{-4}\;\text{GeV}$), while the high densities crossed by atmospheric neutrinos give access to much larger masses ($M_{s} \sim 10^{4}\;\text{GeV}$). Larger values of $M_s$ can be probed in denser environments, as its value increases with the critical density. 

As we are entering the precision era of neutrino physics, the constraints here derived can be significantly improved in the coming years. Through this work, we pave the way for further studies to constrain ST models. In particular, a future supernova detection would allow us to explore larger values of $\lambda$, improved measurements of matter effects in atmospheric neutrinos would further refine the parameter space.\\

\textbf{Acknowledgements --}
The authors thank C.~Burrage, Y.F.~Perez-Gonzalez and E.~Shehu for useful discussions. A.d.G. thanks V. Takhistov and the International Center for Quantum-field Measurement Systems for Studies of the Universe and Particles (QUP/KEK) for their hospitality and the stimulating working environment during which a core part of this work was realised. I.M.S, A.d.G and SSM are supported by the STFC under Grant No.~ST/T001011/1. SSM is additionally supported by funds provided by the Center for Particle
Cosmology at the University of Pennsylvania.

%%%%%%%%%%%%%%%%%%%%%%%%%%%%%%%%%%%%%%%%%%%%%
%%%%%%%%%%%%%%%%%%%%%%%%%%%%%%%%%%%%%%%%%%%%%
% Produces the bibliography via BibTeX.
\bibliographystyle{BiblioStyle}
\bibliography{DraftBiblio}
%%%%%%%%%%%%%%%%%%%%%%%%%%%%%%%%%%%%%%%%%%%%%
%%%%%%%%%%%%%%%%%%%%%%%%%%%%%%%%%%%%%%%%%%%%%
\newpage
\appendix
\begin{center}
    {\textbf{\Large Supplemental Material}}
\end{center}
%%%%%%%%%%%%%%%%%%%%%%%%%%%%%%%%%%%%%%%%%%%%%%%%
%%%%%%%%%%%%%%%%%%%%%%%%%%%%%%%%%%%%%%%%%%%%%%%%
\section{Geodesics and Conserved Quantities in Scalar-Tensor theories}
\label{app:conserved}
In this Section, we derive the geodesics of massive and massless particles in ST both for the Jordan~(JF) and the Einstein~(EF) frames. We then derive integrals of motion as which will then be used to greatly simplify observables.\\

\textbf{Geodesics --}
For the sake of better differentiating between the frames, we will temporarily use tilded quantities to label the JF and untilded ones for the EF.\\

\paragraph{Jordan frame.}
The free massive particle action reads
\begin{equation}
    S=\int\d\tilde\tau~ m \sqrt{\tilde{g}_{\mu\nu}\frac{\d x^\mu}{\d\tilde\tau}\frac{\d x^\nu}{\d\tilde\tau}}~,
\end{equation}
where $\tilde{\tau}$ is the proper time in the Jordan frame, which has to be defined. Now it is the metric the one containing all information about the fifth forces. Minimising this action, we obtain
\begin{equation}
    \label{eq:JF-geo}\frac{\d \tilde{p}^\mu}{\d \tilde\tau} + \frac{1}{m}\tilde{\Gamma}^\mu_{\alpha \beta}\tilde{p}^\alpha \tilde{p}^\beta =\tilde{u}^\alpha \tilde{\nabla}_\alpha \tilde{p}^\mu= 0\,.
\end{equation}
\\ 

\paragraph{Einstein frame.}
The geodesic equation for massive particles comes from minimising the action
\begin{equation}
    S=\int\d\tau~m(\varphi) \sqrt{g_{\mu\nu}\frac{\d x^\mu}{\d\tau}\frac{\d x^\nu}{\d\tau}}~,
\end{equation}
where $m(\varphi)$ allows for the spacetime dependence of the masses, and $\tau$ is the proper time of the particle.
This leads to
\begin{equation}
    \label{eq:EF-geo}\frac{\d p^\mu}{\d \tau} + \frac{1}{m(\varphi)}\Gamma^\mu_{\alpha \beta}p^\alpha p^\beta \Leftrightarrow u^\alpha\nabla_\alpha p^\mu = -\partial^\mu m(\varphi)\,,
\end{equation}
where $p^\mu=m(\varphi)\d x^\mu/\d\tau$. Notice that the EF and JF geodesic equations can be related by redefining the metric and proper time of the particle 
$\d \tilde\tau=A(\varphi)\d\tau$\,.\\

For massless particles, proper time vanishes, so it cannot be used to parametrise the world-line. The go-to choice is taking an affine parameter, which is defined as the parameter that satisfies the geodesic equations. 
In both frames, this leads to the free geodesic equations
\begin{equation}
\begin{split}
    \frac{\d^2 {x}^\mu}{\d \lambda^2} + {\Gamma}^\mu_{\alpha \beta}\frac{\d {x}^\alpha}{\d \lambda} \frac{\d {x}^\beta}{\d \lambda} =&0\,,
\end{split}
\end{equation}
i.e. light-ray geodesics are not affected.\\

%^^^^^^^^^^^^^^^^^^^^^^^^^^^^^^^^^^^^^^^^^^^
\textbf{Conserved Quantity in a Static Background --}
We are interested in deriving the conserved quantity along the geodesics in ST theories. Let us begin by reviewing the conserved quantity in GR in a static background.
In general, if the metric is independent of a coordinate, then it admits a Killing vector. If it is time-independent, the Killing vector reads
\begin{equation}
    \xi^\mu=(1,0,0,0)^\mu\,.
\end{equation}
Then, one can show that the following quantity is conserved along the geodesics:
\begin{equation}
    K=\xi_\mu u^\mu=g_{00} u^0\,,
\end{equation}
where $u^\mu$ is the 4-velocity of the particle. Similarly, we can define the 4-velocity of a stationary observer to be
\begin{align}
    &1=U^\mu U_\mu = (U^0)^2 g_{00}\,, & U^\mu=\frac{1}{\sqrt{g_{00}}} (1,0,0,0)^\mu\,,
\end{align}
and thus
\begin{equation}
    U^\mu=\frac{\xi^\mu}{\sqrt{g_{00}}}\,.
\end{equation}
The energy measured by the static observer is then
\begin{align}\label{eq:omega}
    &\omega=p^\mu U_\mu = \frac{m}{\sqrt{g_{00}}}\, K\,,
\end{align}
where $p^\mu= m u^\mu$.
Therefore, the conserved quantity reads
\begin{align}
    &\label{eq:conserved-naive}K= \frac{\sqrt{g_{00}}\omega}{m}=g_{00}\frac{p^0}{m}\,.
\end{align}
If the mass is constant, one finds the familiar result
\begin{equation}
    \omega_2 =\omega_1 \sqrt{\frac{g_{00}(r_1)}{g_{00}(r_2)}}\,,
\end{equation}
often quoted when studying the Schwartzschild metric.\\

%^^^^^^^^^^^^^^^^^^^^^^^^^^^^^^^^^^^^^^^^^^^

The above argument relies on the use of the geodesic equation, which is nevertheless modified in ST theories in different frames. 
The conservation of $K$ must hold in the Jordan frame as the treatment of the metric did not rely on any assumption on the value og $g_{00}$, besides time-independence. However, it has no reason to be conserved in the Einstein frame due to the modification of the geodesics.
In the following, we will discuss both cases and derive the corresponding conserved quantity in ST theories with a static background.\\

\paragraph{Jordan frame.}
Employing the geodesic equation of Eq.~\eqref{eq:JF-geo}
\begin{equation}
\tilde{u}^\alpha\tilde{\nabla}_\alpha K = \tilde{g}_{00}\tilde{u}^\alpha\tilde{\nabla}_\alpha \tilde u^0=0\,,
\end{equation}
meaning that $K$ as defined in Eq.~\eqref{eq:conserved-naive} is conserved in the Jordan frame.\\

\paragraph{Einstein frame.} 
One can explicitly verify that now the mass-dependent term spoils the conservation of $K$, contrary to what happens in the JF.
Instead, the conserved quantity for a static background is
\begin{equation}
  \label{eq:conserved-real}  K_\text{ST}\equiv m(\varphi) K = g_{00}p^0\,,
\end{equation}
since employing the geodesic equation of Eq.~\eqref{eq:EF-geo} gives
\begin{equation}
    u^\alpha \nabla_\alpha K_\text{ST}=g_{00} u^\alpha \nabla_\alpha p^0=-\partial^0 m(\varphi)=0\,,
\end{equation}
where the last step requires time-independence of the mass. Notice that an analogue derivation can be used to obtain conserved quantities in homogeneous but time-varying backgrounds.\\

All in all, this implies $K_\text{ST}$ is a conserved quantity along the geodesic in both frames.
Let us study how its value changes when switching between the two frames.
We will omit the $_{00}$ subscript and write $m_X$ and $g_X$ with $X=J,E$. The two frames are characterised by
\begin{align}
    &\begin{cases}
        g_J= A^2(x)\,,\\
        m_J=m\,,
    \end{cases} &\begin{cases}
        g_E= 1\,,\\
        m_E= A(x) m\,.
    \end{cases}
\end{align}
Employing the normalisation condition of the four-velocity in the two frames, one can relate the four-velocities
\begin{align}
    &u_X^\mu u_{X,\mu}=1\,, &  u^\mu_E=A(x) u^\mu_J\,.
\end{align}
Employing the definition of $K_\text{ST}$, one finds its value is frame indepedent, i.e.
\begin{equation}
    K_\text{J,ST}=K_\text{E,ST}\,.
\end{equation}

%%%%%%%%%%%%%%%%%%%%%%%%%%%%%%%%%%%%%%%%%%%%%%%%
%%%%%%%%%%%%%%%%%%%%%%%%%%%%%%%%%%%%%%%%%%%%%%%%

\section{Neutrino Observables and Phenomenology}
\label{sec:observables}
We apply the ST framework to neutrino phenomenology. In all case studies, we assume the $\varphi$-background to be static and that neutrinos propagate in a straight line. We employ spherical coordinates centred either at the source or at Earth, so that the metric can be approximated as
\begin{equation}
    g_{\mu\nu}\approx g_{00}dt^2+g_{rr}dr^2-  r^2d\Omega^2\,.
\end{equation}
In this Appendix, we derive the relevant relations in a frame-independent manner. We therefore keep the relevant quantities, such as the metric and the particle parameters (e.g. $G_F$, $m$, and $n_e$), generic throughout the derivation. At any point, the results may then be specialised to either the Einstein~(EF) or Jordan~(JF) frame by using Eqs.~(\ref{eq:frame1})-(\ref{eq:frame2}) and
\begin{align}
    &\text{EF:}~g_{00}=-g_{rr} \approx 1\,, &&\text{JF:}~g_{00}=-g_{rr}\approx A(r)^2\,.
\end{align}
Recall that, for a static background, the following quantity is conserved along the geodesics (cf.~Eq.\eqref{eq:conserved-real})
\begin{equation}
    K_\text{ST}=g_{00}p^0=\sqrt{g_{00}}\omega\equiv E\,,
\end{equation}
where $E$ is the energy measured at the experiment where we impose $g_{00}=1$.
In the following sections, we will work in a generic frame, if not otherwise specified.\\

%&&&&&&&&&&&&&&&&&&&&&&&&&&&&&&&&&&&&&&&&&&&&&&&&&&&&&&&&&&&&&&&&&&&&&&&&&&&&&&&&&&&&&
\textbf{Time delay --}
\label{subsubsec: time delay theory}
To compute the neutrino time delay from Supernova, we need to characterise the motion of a massive particle. By employing the four-velocity normalisation condition
\begin{equation}
    1=g_{\mu\nu}u^\mu u^\nu=g_{00}\left((u^0)^2-(u^r)^2\right)\,,
\end{equation}
and making use of the integral of motion $K_\text{ST}$, one finds
\begin{align}
    \label{eq:four-velocities}&u^r=\sqrt{\frac{K_\text{ST}^2-g_{00}m^2}{(g_{00}m)^2}}\,, &u^0=\frac{K_\text{ST}}{m g_{00}}\,.
\end{align}
In turn, this implies
\begin{align}
    &dt=\left(\frac{K_\text{ST}}{\sqrt{K_\text{ST}^2-g_{00}m^2}}\right)dr\,, 
\end{align}
and thus
\begin{align}
    \label{eq:delay-massive}\Delta t_\nu&= \int\limits_{r_A}^{r_B}dr~\left(\frac{K_\text{ST}}{\sqrt{K_\text{ST}^2-g_{00}m^2}}\right)\\
    &\nonumber=\int\limits_{r_A}^{r_B}dr~\left(\frac{K_\text{ST}}{\sqrt{K_\text{ST}^2-m^2A(r)^2}}\right)\,,
\end{align}
where $r_{A,B}$ are the initial and final positions, respectively. 
Notice that the combination $(g_{00}m^2)$ is also frame-invariant, and the last equality was taken choosing as example the JF. 

It is instructive to compare the result for a massive particle of Eq.~\eqref{eq:delay-massive} to the massless counterpart for the propagation of light. The light-like condition
\begin{equation}
    0=g_{\mu\nu}u^\mu u^\nu=g_{00}\left((u^0)^2-(u^r)^2\right)\,,
\end{equation}
ensures
\begin{align}
\Delta t_\gamma =\Delta r_\gamma\equiv \int\limits_{r_A}^{r_B}dr\,.
\end{align}
Therefore, light rays are not delayed. This is not surprising since conformal invariance of the gauge kinetic term ensures no physical effect can appear. Similarly, the ST impact on neutrinos disappears once $m \to 0$.\\

%&&&&&&&&&&&&&&&&&&&&&&&&&&&&&&&&&&&&&&&&&
\textbf{Flavour Distortion --}
\label{app:flavour}
The time delay introduced by the neutrino mass is only observable over astrophysical distances. However, neutrino mass also produces observable effects at shorter distances through flavour oscillations. In this case, the oscillation length is determined by the mass-squared differences.
In this Section, we investigate the impact of a non-trivial static $\varphi$-background on neutrino oscillations. \\

\paragraph{Vacuum Oscillations.}
Let us begin by reviewing the impact on the vacuum oscillations. 
Denoted by $\ket{\psi}$ the three neutrinos flavour vector, the evolution of the state as a function of the distance can be written as~\cite{Stodolsky:1978ks}
\begin{equation}
    \label{eq:evolution}\ket{\psi(s)}= U\exp\left(i \tilde{S}(s)\right)U^\dagger\ket{\psi(0)}\,,
\end{equation}
where $U$ is the unitary matrix that relates flavour and mass basis $\ket{\psi}=U\ket{\psi_\text{mass}}$, and the action accumulated along the neutrino trajectory in vacuum and in the mass basis ($\tilde{S}(s)$) is given by the integral of the four-momentum ($p^{\mu}$) along the neutrino trajectory.
%the oscillation phase $\tilde{S}(s)$ is given by the integral of the four-momentum ($p^{\mu}$) along the neutrino trajectory.
%
\begin{equation}
    \label{eq:phase}\tilde{S}(s)\equiv  \int_0^s g_{\mu\nu}p^\mu \d x^\nu\,.
\end{equation}

In the following calculations, we will use for the neutrinos the light-ray trajectory to account for the non-measurability of the emission time~\cite{Fornengo:1996ef}
\begin{equation}
    dt=dr\,.
\end{equation}
By employing the expression of the four-velocities of Eq.~\eqref{eq:four-velocities}, the %phase 
action is found to be
\begin{align}
    \tilde{S}(s)&=\int\limits_0^s g_{\mu\nu}p^\mu dx^\nu=\int\limits_0^s \left(K_\text{ST}-\sqrt{K_\text{ST}^2-g_{00}m^2}\right)dr \nonumber\\
    &\approx\frac{1}{2K_\text{ST}}\int\limits_0^s (m^2\,g_{00})\, dr\,,
\end{align}
where we assumed neutrinos are relativistic. This is true for all observables considered in this work, as the EFT consistency ensures that masses can change at most by a factor of two. The result reduces to the known flat result for $g_{00}=1$ and constant $m$. 
The integrand is ST-frame invariant, so we can evaluate it either the JF or the EF. For example, in the JF it reads
\begin{equation}
    \label{eq:S-vacuum}\tilde{S}(s)\approx\frac{m^2}{2E}\int\limits_0^s  dr~A(r)^2\,,
\end{equation}
where we made explicit the identification of the conserved quantity and the measured neutrino energy $E=K_\text{ST}$.\\

%&&&&&&&&&&&&&&&&&&&&&&&&&&&&&&&&&&&&&
\paragraph{Matter Effects.} %yeee
%%%%%%%%%%%%%%%%%%%%%%%%%%%%%%%%%%%%%%%%%%%%%%%%%%%%
We now turn to the derivation of matter effects in a curved background following the discussion of Refs.~\cite{Cardall:1996cd}.
The evolution of the momentum is governed by the equations of motion of the neutrinos, which in the flavour basis are
\begin{equation}
    (i\gamma^\mu\partial_\mu +\gamma^\mu A_{\mu}+ M)\psi=0,
\end{equation}
where $\psi$ is a three-dimensional flavour vector of left-handed neutrinos, $M$ is the mass matrix given by
\begin{equation}
    M=U\left(
        \begin{matrix}
            m_1 & 0 & 0 \\
            0 & m_2 & 0 \\
            0 & 0 & m_3
        \end{matrix}
        \right)U^\dagger,
\end{equation}
and $A^\mu$ encodes the charged-current matter effect
\begin{equation}
    A^\mu=\text{diag}
        \left(-\sqrt{2}G_F N_e^\mu,0,0\right)\,.
\end{equation}
Here, we define the Lorentz invariant number density of electrons as $N_e^\mu=n_e u_e^\mu$,
where the four-velocity for non-relativistic electrons is given by $u_e^\mu\propto(1,0,0,0)$.
Notice that both $G_F$ and $N_e$ are frame-dependent quantities, and as such can carry powers of $A(\varphi)$; their mapping between the two frames can be found in Eqs.~\eqref{eq:frame1}-\eqref{eq:frame2} of the main text.
\begin{figure}[t]
    \centering
    \centering
    \includegraphics[width=\linewidth]{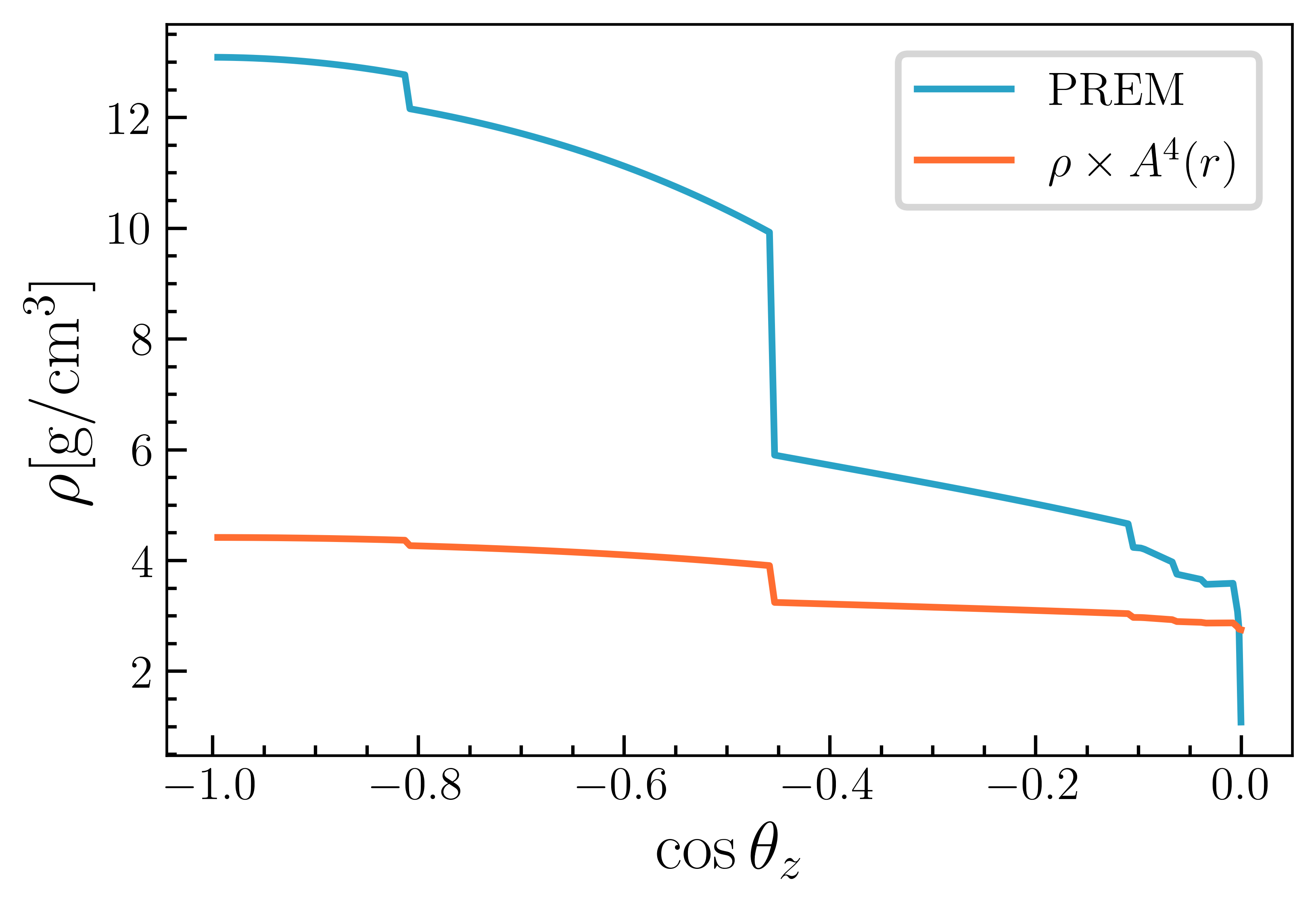}
    \hfill
    \caption{\justifying Earth density profile. In this figure, we compare the Earth's density given by PREM associated with the Jordan frame with the new density obtained in scalar-tensor theories for $n = -0.5$ and $\alpha = 0.5$, computed in the Einstein frame.}
    \label{fig:density}
\end{figure}

In momentum space, the Dirac equation reads
\begin{equation}
    \left[(\slashed{p} +\slashed{A})+ M\right]\psi=0\,,
\end{equation}
where $p^\alpha=\text{diag}(p^\alpha_e,p^\alpha_\mu,p^\alpha_\tau)$ has to be intended as a diagonal matrix in flavour space. In the ultra-relativistic limit, the mass dependence plays a negligible role, so $p^\alpha$ becomes proportional to the identity matrix in flavour space.
Multiplying the above expression from the left by $(\slashed{p} +\slashed{A})- M$ and neglecting terms of $\mathcal{O}(A^2)$ and $\mathcal{O}(AM)$, one finds
\begin{equation}
    g_{\mu\nu}p^\mu \left(p^\nu + 2 A^{\nu}\right)\approx M^2.
\end{equation}
Taking radial coordinates, such that $p^\mu=(p^0,p^r,0,0)$, 
and $g_{rr}=-g_{00}$, we obtain
\begin{equation}
    g_{00}\left({p^{0}}^2-{p^{r}}^2+2p^0 A^0\right)=M^2,
\end{equation}
leading to 
\begin{equation}\label{eq: pr}
   p^{r}\approx p^0-\frac{1}{2 g_{00}p^0}\left(M^2-2g_{00}p^0A^0\right).
\end{equation}
where we are assuming that neutrinos are relativistic and that the potential is smaller than the neutrino energy.
The accumulated action in matter in the flavour basis is given by
%The phase of Eq.~\eqref{eq:phase} can then be written as
\begin{align}
    S(s)&=  \int_0^s g_{00}\left(p_0 dt -p^r dr\right)\,,\\
    &\nonumber \approx \int_0^sdr~ \frac{g_{00}}{2p^0 g_{00}}\left(M^2-2g_{00}p^0A^0\right)\,,
\end{align}
where $\tilde S(s)=U^\dagger S(s)U$.
The result can be conveniently written by making explicit the conserved quantity of Eq.~\eqref{eq:conserved-real}, which can be identified with the measured energy, $K_\text{ST}=E$,
\begin{equation}
    S(s)\approx \int_0^sdr~ g_{00}\left(\frac{M^2}{{2E}}+V_\text{CC}\right)\,,
\end{equation}
where $(V_\text{CC})_{ij}=\sqrt{2}n_e G_Fu_e^0\delta_{ij,e}$, and we kept $u_e^0$ explicit for convenience to more easily compare between the Jordan and the Einstein frames. One can show that $g_{00}V_\text{CC}$, and therefore $S(s)$, is frame invariant.
For example, in the JF it reads
\begin{equation}\label{eq:osc}
    S(s)\approx \int_0^sdr~ A(r)^2\left(\frac{M^2}{{2E}}+V_\text{CC}\right)\,.
\end{equation}
As can be seen, the vacuum part and the matter effects are equally scaled. 
%%%%%%%%%%%%%%%%%%%%%%%%%%%%%%
%%%%%%%%%%%%%%%%%%%%%%%%%%%%%%
\section{Neutrino Oscillations}
\label{app:oscillations}

Scalar–tensor (ST) theories can influence neutrino flavour oscillations by modifying neutrino masses and, consequently, the oscillation length, as discussed in previous sections. When neutrinos propagate through matter, these theories also affect the matter potential. In this work, we focus on the flavour oscillations of atmospheric neutrinos.

\begin{figure*}
\includegraphics[width=\linewidth]{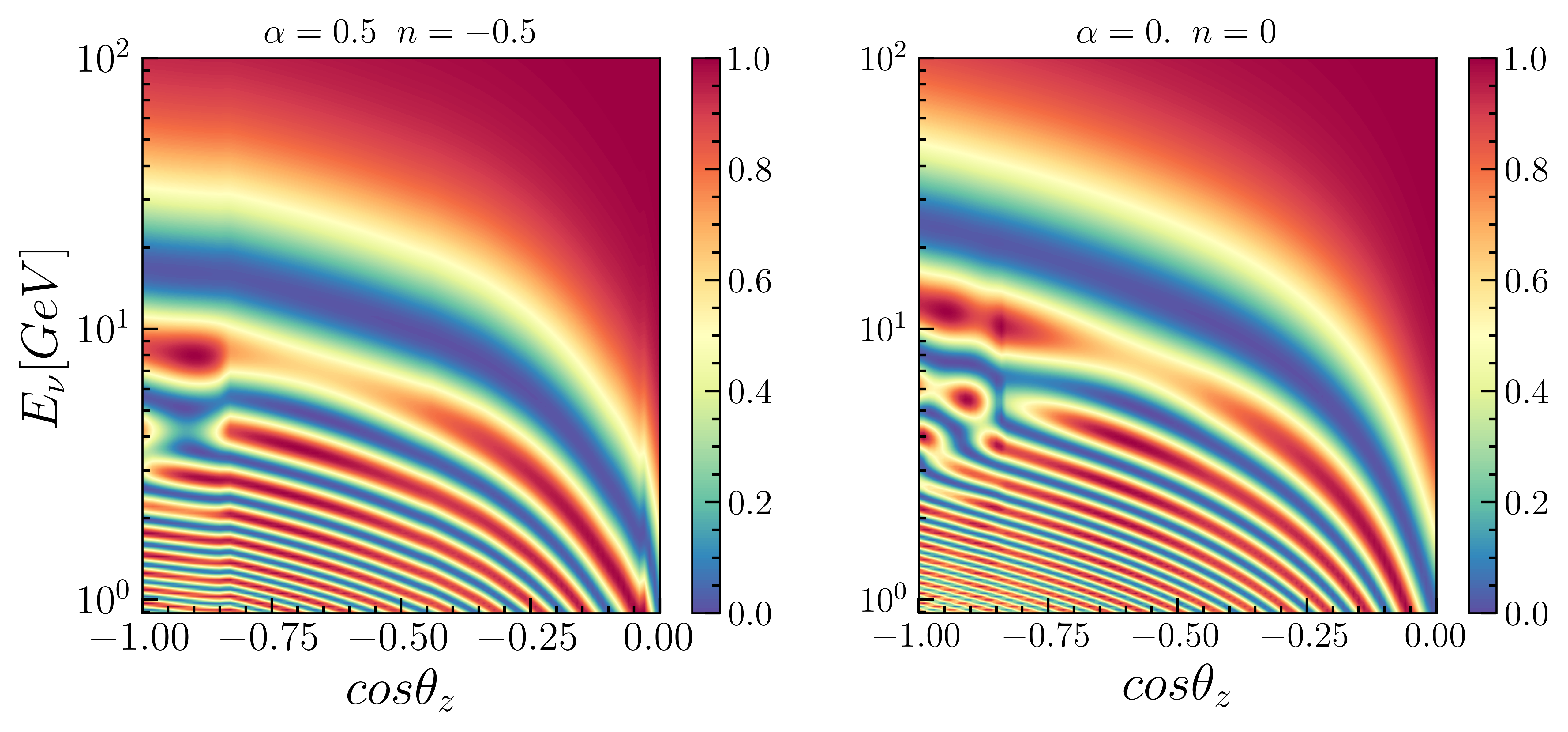}
\captionof{figure}{\justifying Muon survival probability ($P(\nu_{\mu} \rightarrow \nu_{\mu})$). We calculate the probability that muon neutrinos survive after crossing the Earth, for both the standard scenario (right) and the ST case with parameters 
$ n = -0.5 $ and $ \alpha = 0.5 $ (left).}
\label{fig:Oscillo}
\end{figure*}

Before reaching the detector, atmospheric neutrinos travel through the Earth. For our analysis, we adopt the PREM model as the reference density distribution. In the Einstein frame, neutrino masses inside the Earth differ from those measured at the surface, and as consequence, the effective oscillation length differs from what is naively expected. For illustration purposes, in vacuum, the flavour evolution is dictated by (cfr.~Eqs.~\eqref{eq:evolution}-\eqref{eq:S-vacuum})
\begin{equation}
    \ket{\psi(L)}\sim e^{i \left(\frac{L}{ L^{\text{eff}}_{\text{osc}}}\right)}\ket{\psi(0)}\,,
\end{equation}
where we identified an effective oscillation length
\begin{equation}
    L^{\text{eff}}_{\text{osc}} = \frac{4 E L}{\Delta m^2 \int dr\, A^2(r)}\,.
\end{equation}
%
% \red{This equation is intended only for a qualitative description, since the effective neutrino masses also vary at each point along the neutrino trajectory~\cite{Denton:2016wmg}.}\adgC{Isn't it already encoded in $A(r)$?}
For \( A(r) \neq 1 \), the oscillation length can be either suppressed or enhanced. Using the parametrisation in Eq.~\cref{eq:effective}, Fig.~\ref{fig:density} illustrates how the density changes for \( n = -0.5 \) and \( \alpha = 0.5 \) in the Einstein frame. For these values, we find \( A^2 \simeq 0.8 \) at \( \rho = 13~\text{g/cm}^3 \), which increases the oscillation length from approximately the Earth’s diameter at \( E \sim 20~\text{GeV} \) to about \( 1.8 \times 10^{4}~\text{km} \). This shift lowers the energy corresponding to the first oscillation minimum, as shown in Fig.~\ref{fig:Oscillo}.

Beyond changes in oscillation length, ST theories also modify matter effects by altering the density profile (see Fig.~\ref{fig:density}). In regions of constant density or under adiabatic evolution, the factor \( A(r) \) acts as a common multiplier in both the mass and potential terms along the entire trajectory, leaving the mixing angles unchanged. Consequently, no observable effect is expected when oscillation probabilities depend only on mixing angles, as in the Sun or supernovae. 
To illustrate this, we use Eq.~\ref{eq:amplitude} for a two-neutrino flavour system. In this case, the Hamiltonian becomes
\begin{equation}
    H_{2\nu} = \frac{1}{2E}\left(\begin{array}{cc}
        -\Delta m^2 \cos 2\theta + 2E V_{CC} & \Delta m^2 \sin 2\theta \\
        \Delta m^2 \cos 2\theta  & \Delta m^2 \cos 2\theta - 2EV_{CC}
    \end{array}\right)
\end{equation}
This Hamiltonian is diagonalised by a rotation matrix that depends on the matter potential, \(\tilde{U}(\tilde{\theta})\). We can expand this unitary matrix as

\begin{equation}
\tilde{U}(\tilde{\theta}) \approx \tilde{U}(\tilde{\theta_0}) + \frac{1}{2}\frac{d\tilde{U}}{dr},
\end{equation}
which, to leading order, remains unitary (\(\tilde{U}\tilde{U}^{\dagger} = 1\)). Using this expansion to diagonalise the Hamiltonian, we have
\begin{align}
   H_{2\nu} &= \frac{1}{2E}\tilde{U}^{\dagger}(\theta)\text{diag}(-\Delta\tilde{m}^2, \Delta\tilde{m}^2)\tilde{U}(\theta)\nonumber\\
   &=\frac{1}{2E}\left(\tilde{U}^{\dagger} (\theta_{0}) + \frac{1}{2}\frac{d\tilde{U}^{\dagger}}{dr}\right)\text{diag}(-\Delta\tilde{m}^2, \Delta\tilde{m}^2)\times\nonumber\\
   &\qquad\times\left(\tilde{U} (\theta_{0}) + \frac{1}{2}\frac{d\tilde{U}}{dr}\right)
\end{align}
After rearranging terms, this becomes
\begin{equation}
H_{2\nu} = [H^{0}_{2\nu}, \mathcal{O}(d\tilde{\theta}/dr)],
\end{equation}
where
\begin{equation}
H^{0}_{2\nu} = \frac{1}{2E}\tilde{U}^{\dagger}(\theta_0)\,\text{diag}(-\Delta\tilde{m}^2, \Delta\tilde{m}^2)\,\tilde{U}(\theta_0),
\end{equation}
and
\begin{equation}
\mathcal{O}(d\tilde{\theta}/dr) = \begin{pmatrix}
1 & -d\tilde{\theta}/dr \\
d\tilde{\theta}/dr & 1
\end{pmatrix},
\end{equation}
where
\begin{equation}
d\tilde{\theta}/dr = \frac{1}{2}\frac{dV_{CC}/dr \,\sin 2\theta}{(\cos 2\theta - 2E V_{CC})^2}.
\end{equation}
The evolution is adiabatic if \( d\tilde{\theta}/dr < 1 \), and this condition is independent of \( A(r) \). Therefore, we can conclude that ST does not modify neutrino evolution in the Sun or in supernovae.

\begin{figure*}[t]
\includegraphics[width=\linewidth]{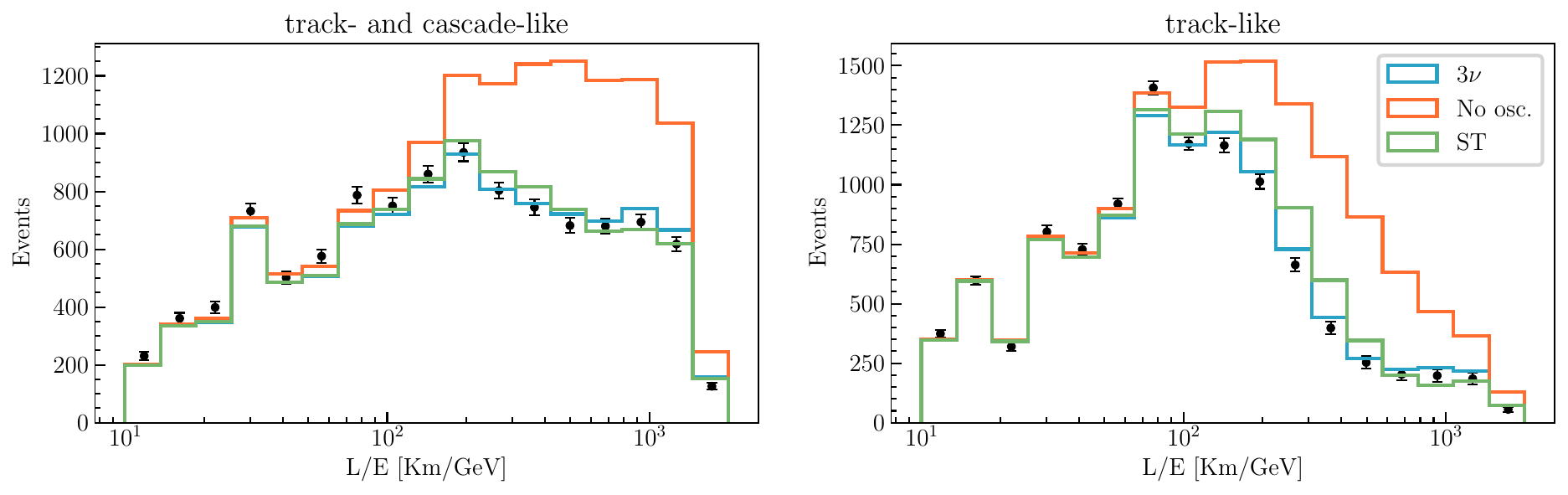}
\captionof{figure}{\justifying Event distribution. We present data corresponding to $9.3$ years of IceCube DeepCore observations (black dots), binned according to the $L/E$ ratio and including statistical uncertainties. The distributions are shown for track- and cascade-like events (left) and for track-like events only (right). Alongside the data, we display the MC predictions for three scenarios: the standard $3\nu$ scenario, a no-flavour-oscillation case, and the ST scenario with parameters $n = -0.5$ and $\alpha = 0.5$}
\label{fig:Events}
\end{figure*}

The situation differs for the Earth, where the evolution is non-adiabatic. In this case, ST theories affect oscillations when neutrinos traverse resonant regions, as illustrated in Fig.~\ref{fig:Oscillo} for energies around \(10~\text{GeV}\) and trajectories crossing the entire Earth. In Fig.~\ref{fig:Events}, we show the impact of ST on the event distribution. We follow the analysis of $9.3$~years of IceCube DeepCore data, using two data samples: track- and cascade-like events (left) and track-like events (right). The data, including statistical uncertainties, are shown as black dots. In the same figure, we also include the predictions for the standard $3\nu$ scenario with \(\Delta m^2_{31} = 2.5\times 10^{-3}\;\text{eV}^2\) and \(\sin^2\theta_{23} = 0.51\), as well as the prediction for the no-oscillation case. For the ST prediction, we use \(n = -0.5\) and \(\alpha = 0.5\), following the parametrisation given in Eq.~\ref{eq:effective}. As shown, the largest impact occurs in the track sample for baselines crossing the Earth and energies around \(25~\text{GeV}\), corresponding to \(L/E \sim 400\;\text{km/GeV}\). In Fig.~\ref{fig:Events}, systematic uncertainties are not included; however, the full analysis incorporates all systematics as described in Ref.~\cite{IceCubeCollaboration:2024ssx}.

Atmospheric neutrino oscillations above the GeV energy region are mainly sensitive to $\Delta m^2_{31}$ and $\sin^2\theta_{23}$. For each value of $n$, we have analyzed IceCube data considering three parameters: $\Delta m^2_{31}$, $\sin^2\theta_{23}$, and $\alpha$. The bounds on $\alpha$ are obtained after profiling over $\Delta m^2_{31}$ and $\sin^2\theta_{23}$, without imposing any external constraints on these two parameters.

\begin{figure*}[t]
\includegraphics[width=0.8\linewidth]{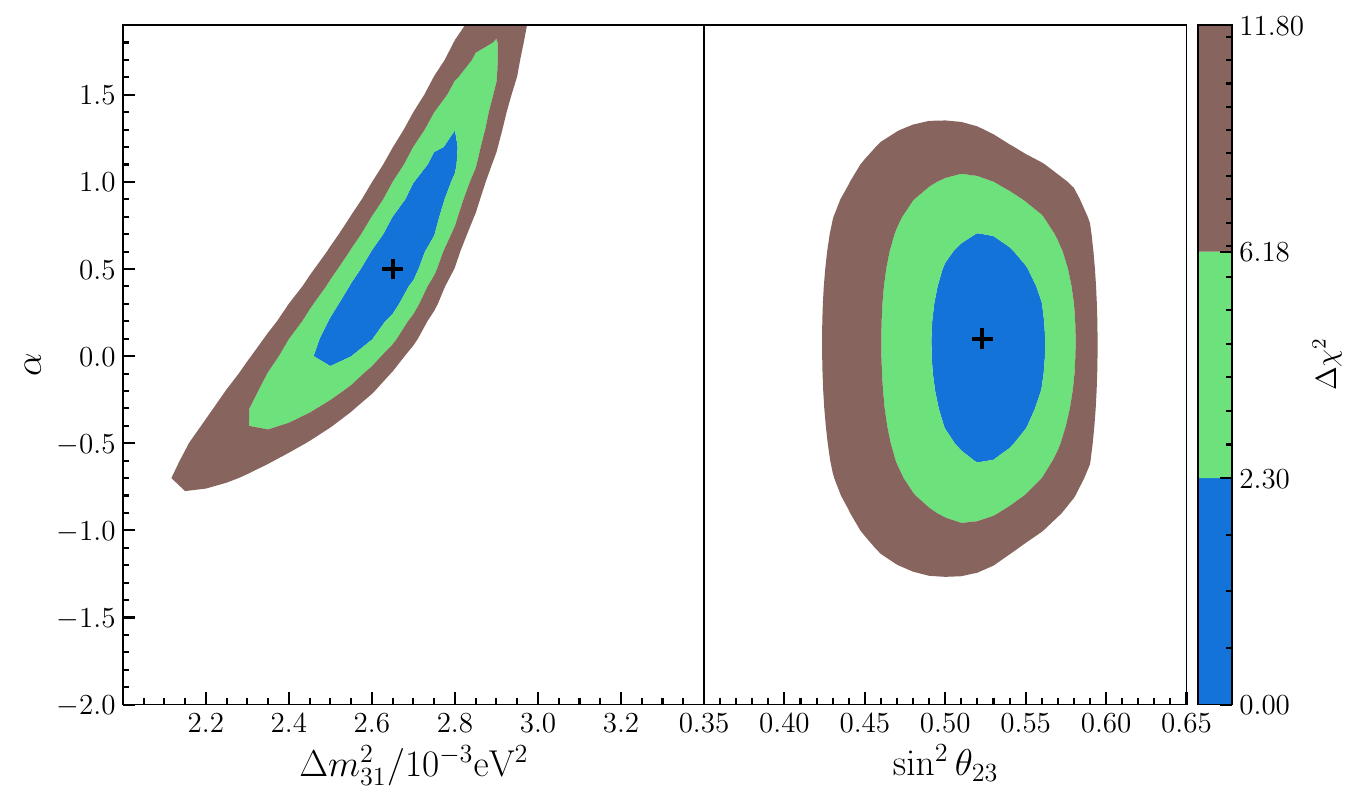}
\captionof{figure}{\justifying Allowed regions for $\alpha$, $\Delta m^2_{31}$, and $\sin^2\theta_{23}$ for $n = -0.5$. The $\chi^2$ has been minimised over the parameter not shown in each panel. The $1\sigma$, $2\sigma$, and $3\sigma$ allowed regions are shown, corresponding to $\Delta \chi^2 = 2.3$, $6.18$, and $11.83$, respectively.}
\label{fig:correlation}
\end{figure*}

In Fig.~\ref{fig:correlation}, we show the $1\sigma$, $2\sigma$ and $3\sigma$ allowed regions for two parameters after minimising over the third one for $n = -0.5$. As can be seen, there is a correlation between $\Delta m^2_{31}$ and $\alpha$. This behaviour can be understood from Eq.~\ref{eq:osc}, since the oscillation phase acquired by neutrinos along their trajectory is multiplied by $A(r)^2$. For $n = -0.5$, $A(r)^2$ is smaller than unity, which can be compensated by larger values of $\Delta m^2_{31}$.

\end{document}